\documentclass{nature}

\usepackage[nomargin,inline,draft,author=]{fixme}
\usepackage{soul}
\fxusetheme{color}
\definecolor{fxwarning}{rgb}{0.8,0.0000,0.0000}

\usepackage{graphicx}

\usepackage[font=small,labelfont=bf]{caption}

\spacing{2}

\usepackage{amsmath}

\usepackage{bm}

\usepackage{xr}
\makeatletter
\newcommand*{\addFileDependency}[1]{
  \typeout{(#1)}
  \@addtofilelist{#1}
  \IfFileExists{#1}{}{\typeout{No file #1.}}
}
\makeatother

\newcommand*{\myexternaldocument}[1]{
    \externaldocument{#1}
    \addFileDependency{#1.tex}
    \addFileDependency{#1.aux}
}

\myexternaldocument{supplementary_revision}

\title{Silicon photonic-electronic neural network for fibre nonlinearity compensation}

\author{Chaoran Huang$^{1,4}$,~Shinsuke Fujisawa$^2$,~Thomas Ferreira de Lima$^{1}$,~Alexander N. Tait$^{1}$,~Eric C. Blow$^1$~,Yue Tian$^{2}$,~Simon Bilodeau$^1$,~Aashu Jha$^1$,~Fatih Yaman$^{2}$,~Hsuan-Tung Peng$^1$,~Hussam G. Batshon$^{2}$,~Bhavin J. Shastri$^{1,3}$,~Yoshihisa Inada$^{5}$,~Ting Wang$^{2}$,~and Paul R. Prucnal$^1$}

\begin{document}

\maketitle

\begin{affiliations}
 \item Department of Electrical Engineering, Princeton University, Princeton, NJ 08542, USA
 \item NEC Laboratories America, Inc., 4 Independence Way, Princeton, NJ 08540, USA
 \item Department of Physics, Engineering Physics and Astronomy, Queen’s University, Kingston, ON K7L 3N6, Canada
 \item Department of Electronic Engineering, the Chinese University of Hong Kong
 \item Submarine Network Division, NEC Corporation, 108-8001 Tokyo, Japan
\end{affiliations}


\begin{abstract}

In optical communication systems, fibre nonlinearity is the major obstacle in increasing the transmission capacity. Typically, digital signal processing techniques and hardware are used to deal with optical communication signals, but increasing speed and computational complexity create challenges for such approaches. Highly parallel, ultrafast neural networks using photonic devices have the potential to ease the requirements placed on the digital signal processing circuits by processing the optical signals in the analogue domain. Here we report a silicon photonic–electronic neural network for solving fibre nonlinearity compensation of submarine optical fibre transmission systems. Our approach uses a photonic neural network based on wavelength-division multiplexing built on a CMOS-compatible silicon photonic platform. We show that the platform can be used to compensate optical fibre nonlinearities and improve the signal quality (Q)-factor in a 10,080 km submarine fibre communication system. The Q-factor improvement is comparable to that of a software-based neural network implemented on a 32-bit graphic processing unit assisted workstation. Our reconfigurable photonic-electronic integrated neural network promises to address pressing challenges in high-speed intelligent signal processing. 

\end{abstract}


Artificial Neural networks (ANNs) can learn from examples and perform tasks without being programmed with explicit rules, providing a powerful tool to solve problems in many disciplines\cite{najafabadi2015deep,deng2014deep}. Current applications with ANNs, such as image and voice recognition, are typically performed locally at kHz speed or on cloud servers -- which high latency -- using conventional computers. But many emerging ANN applications demand gigahertz bandwidth, real-time operations, and low power consumption~\cite{de2019machine,wang2009fast,shi2019deep}. ANNs have, for example, been applied in optical fiber communications where they can learn optical fiber transmission models from received signals without prior knowledge of transmission conditions\cite{zibar2015machine,khan2019optical}. The ANNs can be used to compensate fibre nonlinearity impairment, the major limiting effect of increasing capacities in long-distance transmission systems\cite{zhang2019field}. Compared to conventional fiber nonlinearity compensation approaches, such as digital backpropagation (DBP)\cite{ip2008compensation}, an ANN can achieve comparable signal quality improvement with lower computational complexity\cite{giacoumidis2020real}. However, such ANNs have only been implemented in software, and thus the data processing was not in real-time\cite{zhang2019field}.

In optical communication systems, the signal bandwidth is tens of gigahertz, and the signals must be processed in real-time. Bounded by the clock rate of digital electronic hardware (typically application-specific integrated circuits; ASICs), processing such high-speed signals requires massive parallelisation. As a result, the circuit complexity scales with the computational complexity of the digital signal processing (DSP) algorithms. At the same time, the signal processing unit needs to be built into a pluggable transceiver module and its power consumption needs to fit into the power envelope of network switches, typically tens of watts including both optical and electrical modules\cite{morero2016design}. The challenges in high circuit complexity with a tight power budget have prohibited implementing high-performance, but computationally intensive, ANN based algorithms\cite{du2014digital}. And efforts have been focused on developing algorithms that lead to a compromise between transmission link performance and DSP complexity.


An alternative approach is to explore hardware that intrinsically offers both high bandwidth and low power consumption in ANN applications, and photonic neural networks (PNNs) could potentially provide such capabilites\cite{prucnal2017neuromorphic,shastri2021photonics}. PNNs aim to leverage high-speed optical devices to mimic the essential computing primitives (neurons and synapses) and connect them into a neural network with highly parallel and dense optical interconnects. As a result, PNNs are potentially capable of supporting real-time ANN implementations with tens of gigahertz bandwidth in a single pipeline. The approach could thus address DSP related hardware constraints, such as analogue-to-digital performance and computational complexity, and help cope with the increasing data transmission capacity.

PNNs have previously been demonstrated on various platforms, including free-space\cite{brunner2015reconfigurable,zuo2019all}, optical fibres\cite{hill2002all,fok2011signal,brunner2013parallel}, diffractive optics\cite{lin2018all,bueno2018reinforcement}, and integrated photonic circuits\cite{tait2017neuromorphic,peng2018neuromorphic,shen2017deep,feldmann2019all,vandoorne2014experimental}. The past decade has seen a rapid development of silicon photonics, driven by the rising demands in optical communication capacities. These advances have provided fast and efficient optical modulators and detectors for optical communications, and have led to high-density integrated optoelectronic devices and interconnects that can perform scalable information processing and computing tasks. However, many of the integrated devices demonstrated so far have focused on building analogue interconnects for multiply-accumulate (MAC) computations for linear matrix-vector multiplication operations~\cite{lin2018all,shen2017deep}with nonlinear activation functions (neurons) implemented offline (in the digital domain) or with slow optics. Alternatively, there have been demonstrations of neurons with reconfigurable activation functions but without co-integrated linear operations~\cite{tait2019silicon,huang2020chip}. The lack of a fully integrated neuron with network compatibility prevents PNNs from performing real-time processing with gigahertz bandwidth, which is vital for many signal processing applications.

In this Article, we report an integrated silicon PNN system that integrates a full NN model with neurons and synapses, and demonstrate for the first time that a PNN is capable of simultaneously providing accurate weighting, summation, and biased nonlinearity (i.e., neurons) for high-speed signals. Integrating fast photonic neurons on chip is a critical step toward achieving high-speed NNs for real-time intelligent signal processing. Our PNN is a reconfigurable photonic-electronic integrated circuit and is programmable to perform different tasks. We demonstrate a unique application of PNN for fibre nonlinearity compensation (NLC) in submarine optical fibre communications transmission systems. ANN parameters trained on a computer in advance are accurately uploaded to our photonic chip, which can then be used to perform inference acceleration tasks. Our platform can predict optical fibre nonlinearities and improve the signal quality (Q)-factor in a 10,080 km submarine fibre communication system. The Q-factor improvement is comparable to the results obtained from a software-based neural network running on a 32-bit graphics processing unit (GPU)-assisted workstation. The results illustrate that our photonic analogue computing circuit can preserve signal integrity, while offering potential speed and gigahertz bandwidth advantages without sacrificing accuracy. The NLC application could also be generalized as a regression problem, and extended to solve different real-world application problems currently unreachable with conventional computing technologies.

\section{Neural network model for fiber nonlinearity compensation}


We investigate a 10,080 km submarine optical transmission system carrying a single channel Gbaud polarization-multiplexed (PM)-16 quadrature amplitude modulation (QAM) signal. While recent demonstrations have shown the transmission capacities have approached the Shannon limit in the linear regime of optical fibers, the nonlinear impairment remains the major limiting effect in long-distance transmission systems~\cite{essiambre2010capacity}. The evolution of the signals in the fiber link is described by the Nonlinear Schr\"odinger Equation:
\begin{equation}\label{eq: nlse}
\frac{\partial{u_{x/y}(t,z)}}{\partial{z}} + \underbrace{i\frac{\beta_{2}}{2}\frac{\partial^2{u_{x/y}(t,z)}}{\partial^2{t}}}_\text{\footnotesize dispersion}=\underbrace{i\frac{8}{9}\gamma[\mid u_{x/y}(t,z)\mid^2 + \mid u_{y/x}(t,z)\mid^2 ]\mid u_{x/y}(t,z)}_\text{\footnotesize Kerr nonlinearity},
\end{equation}
where $u_{x/y}(t,z)$ is the optical field at the x and y polarizations, respectively, $\beta_2$ is the group velocity dispersion governing the linear impairment, and $\gamma$ is the Kerr nonlinear coefficient governing the nonlinear impairment in the fiber transmission systems. Although state-of-the-art DSP can overcome linear impairments, nonlinear impairments remain difficult to overcome due to the difficulty in applying computationally expensive nonlinear compensation algorithms using ASICs. The nonlinearity impairment can be treated as the the perturbation due to the nonlinear effects, \emph{i.e.}, $u_{x/y}(t,z) = u_{0, x/y}(t, z) + \Delta u_{x/y}(t,z)$, where $u_{0, x/y}(t, z)$ is the solution of linear propagation and $\Delta u_{x/y}(t,z)$ is the nonlinear perturbation. The idea of NLC is to build up a model of the nonlinear perturbation $\Delta u_{x/y}(t,z)$ with the abundance of transmission data using a NN, and remove the distortion by subtracting the perturbation from the received optical field $u_{x/y}(t,z)$.

Here we first consider how fiber nonlinearities impair the signal on the x-polarization as an example. We define a term \textit{triplet} as $T = H_{n}H_{m+n}^* H_{m} + V_{n}V_{m+n}^* H_{m}$, where $H$ and $V$ are the received symbol sequences at the x- and y- polarizations, respectively. $m$ and $n$ are symbol indices with respect to the symbols of interest $H_0$. Triplet $T$ represents the nonlinearities added on the X-polarization signals, which is caused by the nonlinear interactions among the signals at the same polarization (i.e., $H_{n}H_{m+n}^* H_{m}$) and orthogonal polarizations ($V_{n}V_{m+n}^* H_{m}$), as shown in Eq.~\ref{eq: triplet}. Triplet $T$ corresponds to the nonlinear terms on the right side of Eq.~\ref{eq: nlse}. Then, the nonlinear perturbation can be approximated by a combination of triplets $T$~\cite{tao2011multiplier}:
\begin{equation}\label{eq: triplet}
\begin{split}
\Delta u_{x/y}(0,z) = \sum P_0^{3/2}(\underbrace{H_{n}H_{m+n}^* H_{m}}_{\begin{subarray}{c}\text{\footnotesize intra-polarization}\\ \text{\footnotesize nonlinearity}\end{subarray}}
+ \underbrace{V_{n}V_{m+n}^* H_{m})}_{\begin{subarray}{c}\text{\footnotesize inter-polarization}\\ \text{\footnotesize nonlinearity}\end{subarray}}     C_{m,n}
\end{split}
\end{equation}
where $P_0$ is the launch power, $C_{m,n}$ are the nonlinear perturbation coefficients.




\begin{figure}
\includegraphics[width=1\linewidth]{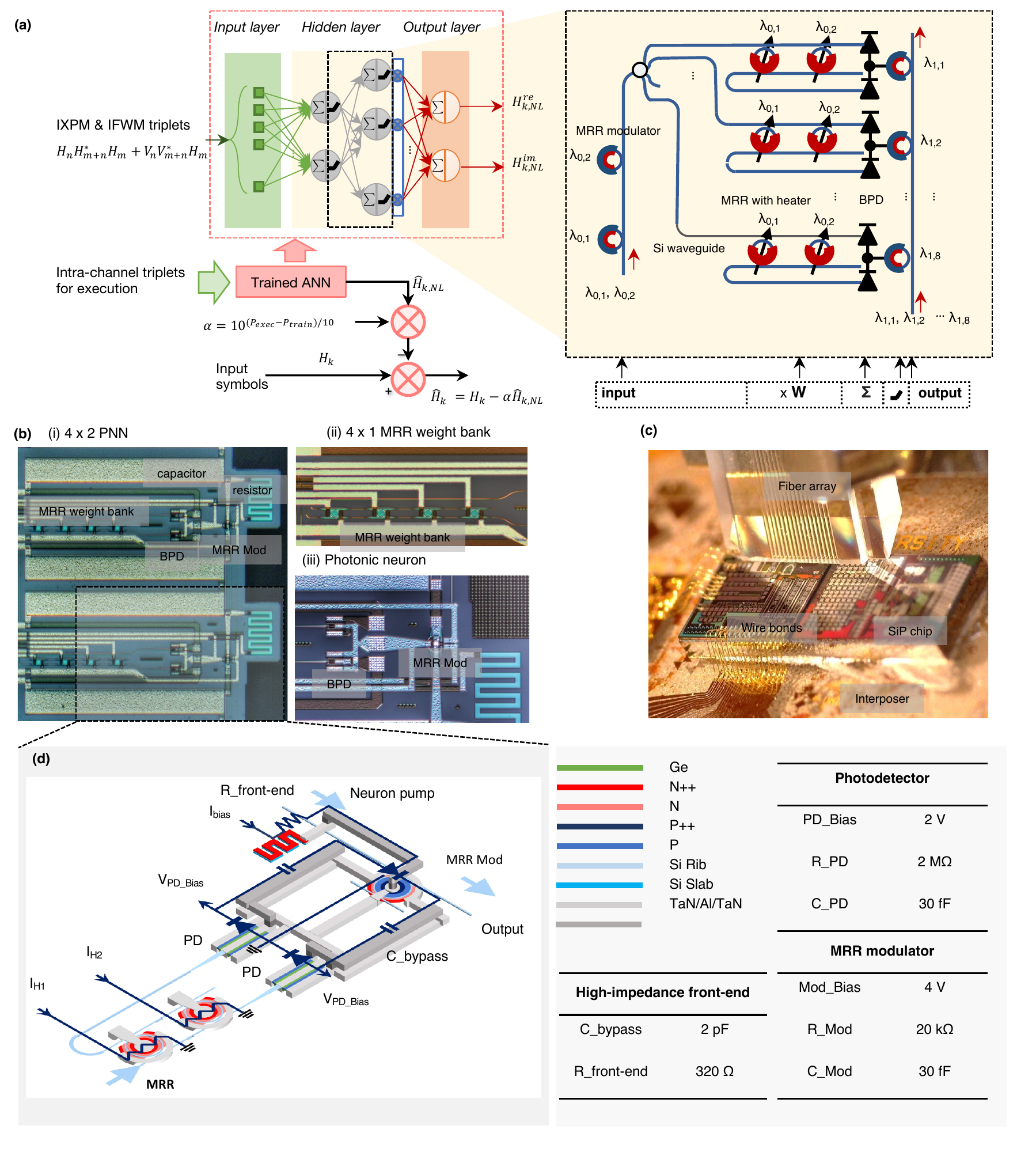}
\spacing{1.}
\vspace*{-10mm}
\caption{\label{fig:system} (a)~(left) Neural network (NN) model for fiber nonlinearity compensation (NLC). The NN has an input layer with nonlinear triplets, 2 hidden layers with of 2 and 8 neurons in each layer, and an output layer with 2 output nodes corresponding to the real and imaginary parts of the estimated nonlinearity on one polarization, and (right) PNN implementation of the second hidden layer of the NN model based on weight-and-broadcast architecture\cite{tait2014broadcast,tait2017neuromorphic}. Red (blue) arcs are n-doped (p-doped) regions. (b) A false-color confocal micrograph of the PNN device. (c) Chip packaging and optical coupling setup. (d) A detailed circuit diagram of a photonic neuron.}
\end{figure}

Fig.~\ref{fig:system}(a) shows the NN model for NLC. The NN is a fully connected feed-forward network consisting of an input layer with 892 triplets, two hidden layers with 2 and 8 neurons, respectively, and an output layer. The model is trained using the signal on either x or y polarization only, but the trained parameters can be applied to both polarizations. Here, we choose polarization x as an example. The inputs of the NN are the triplets $T$ for the x-polarization signal~\cite{zhang2019field}.
The two outputs of the NN are the real and imaginary components of the estimated nonlinearity $\hat{H}_{k, NL}$ at the k-th symbol. The recovered symbol $\hat{H}_{k}$ is obtained by subtracting $\hat{H}_{k,~NL}$ from the received symbol $H_{k}$, i.e. $\hat{H}_{k}=H_{k}-\hat{H}_{k,~NL}$. The NN is trained by searching for the tensor parameters that minimize the mean squared error (MSE) between the $\hat{H}_{k}$ and the transmitted symbol $A_{x,k}$. After training, the NN can execute the inference task of estimating the optical nonlinearity of every incoming symbol. The NN is much simpler than those used for many other applications (e.g. image processing). Nevertheless, high-speed communications signals require fast operating speed, posing many challenges to today's digital electronic hardware. For example, to process 32 Gbaud PM-16QAM signals, NN hardware must have a throughput of approximately 232.4 TOPS. A typical NN accelerator, Tensor Processing Unit (TPU), offers a peak operation throughput of 92 TOPS\cite{jouppi2017datacenter}. As a result, the need for evaluating NN models for optical communication signals necessitates the development of new hardware. 

\section{Photonic hardware implementation}
The NN model are implemented using a PNN with WDM-based broadcast-and-weight architecture, as shown in the yellow region of Fig.~\ref{fig:system}(a). The information of each neuron is encoded on a unique wavelength of light and multiplexed on a single optical waveguide.
In the feed-forward network, the neuron output in the previous layer is first broadcast to all neurons in the next layer by a power splitter, then weighted in parallel with an array of microring resonators (MRR) in an MRR weight bank configuration, and finally summed by a balanced photodetector.  Weights are applied by the MRR banks by tuning the partial transmission of signals. The weighted signals are detected, generating photocurrent to drive the optical modulators in the second layer. This process involves a nonlinear conversion of optical power into electrical current and back into the signal pathway, and hence serves as the nonlinear activation function (i.e. neuron) in the PNN\cite{tait2019silicon}. 



The signal pathways in the PNN can be fabricated using standard silicon photonic platforms, which provides critical aspects of feasibility and economies of scaling to the full NN model for NLC. Several notable studies have demonstrated the integration of thousands of photonic components on a single silicon photonic chip\cite{sun2013large,poulton20208192}. 
For the first time, we show that the PNN can provide accurate and fast weighting, summation, and biased nonlinearity for high-speed signals, and we experimentally implement the second hidden layer of the NN model using a 4$\times$2 PNN, as shown in Fig.~\ref{fig:system}(b). In particular, integrating fast photonic neurons on chip is a critical move toward achieving high-speed NNs for real-time applications such as NLC. In the Discussion section, we discuss that the PNN can be robustly scaled to incorporate the full NN model.

The PNN device comprises two arrays of MRR weight banks connected to two photonic neurons. The second hidden layer consists of 8 neurons, and each neuron is connected to the two neurons in the first hidden layer by a 2$\times$2 weight matrix. To emulate the second hidden layer with the 4$\times$2 PNN circuit, we feed the two first-layer outputs to the PNN four times with four subsets of weights and biases.
The photonic neuron circuit diagram is shown in Fig.~\ref{fig:system}(d). The MRR weight bank is implemented with in-ring N-doped photoconductive heaters\cite{tait2018feedback} and provides the key functionality to configure connection strengths (\emph{i.e.} weights or synapses). The weight is determined by the MRR transmission, which can be thermally tuned by adjusting the electrical current applied to the N-doped heaters\cite{tait2018feedback}. The N-doped heaters allow continuous, multi-channel control of the MRR weight bank with accuracies up to 8 bits\cite{huang2020demonstration,zhang2021microring}, which quantifies the precision of weights implementation affected by the noise in the setup. The bit resolution obtained in photonic matrix multiplier is comparable that used in DSP ASICs. The MRR weight bank has two complementary optical outputs, each of which is detected by a germanium-on-silicon photodetector.
The two photodetectors form a balanced photodetector~\cite{hai201316}, where the output photocurrent represents the subtraction operation between the two MRR weight bank outputs, resulting in a complementary –1 to +1 continuous weight range. The photocurrent, combined with a forward bias current $I_{bias}$, modulates the transmission of the MRR modulator (\emph{i.e.} photonic neuron) via free-carrier injection to the p-n junction, and hence modulates the optical power of a continuous-wave laser (labeled as "neuron pump"). The on-chip capacitor and resistor provide a network matching circuit for efficient optical-electrical-optical (OEO) conversion (see Supplementary information \textit{Photonic neuron simulation} for detailed circuit analysis). The MRR modulator exhibits nonlinear electrical-to-optical transfer functions, which produce the activation function in the NN. Furthermore, the neuron biases can be configured by adjusting the electrical current applied to the MRR modulator.

\begin{figure}
\includegraphics[width=1\linewidth]{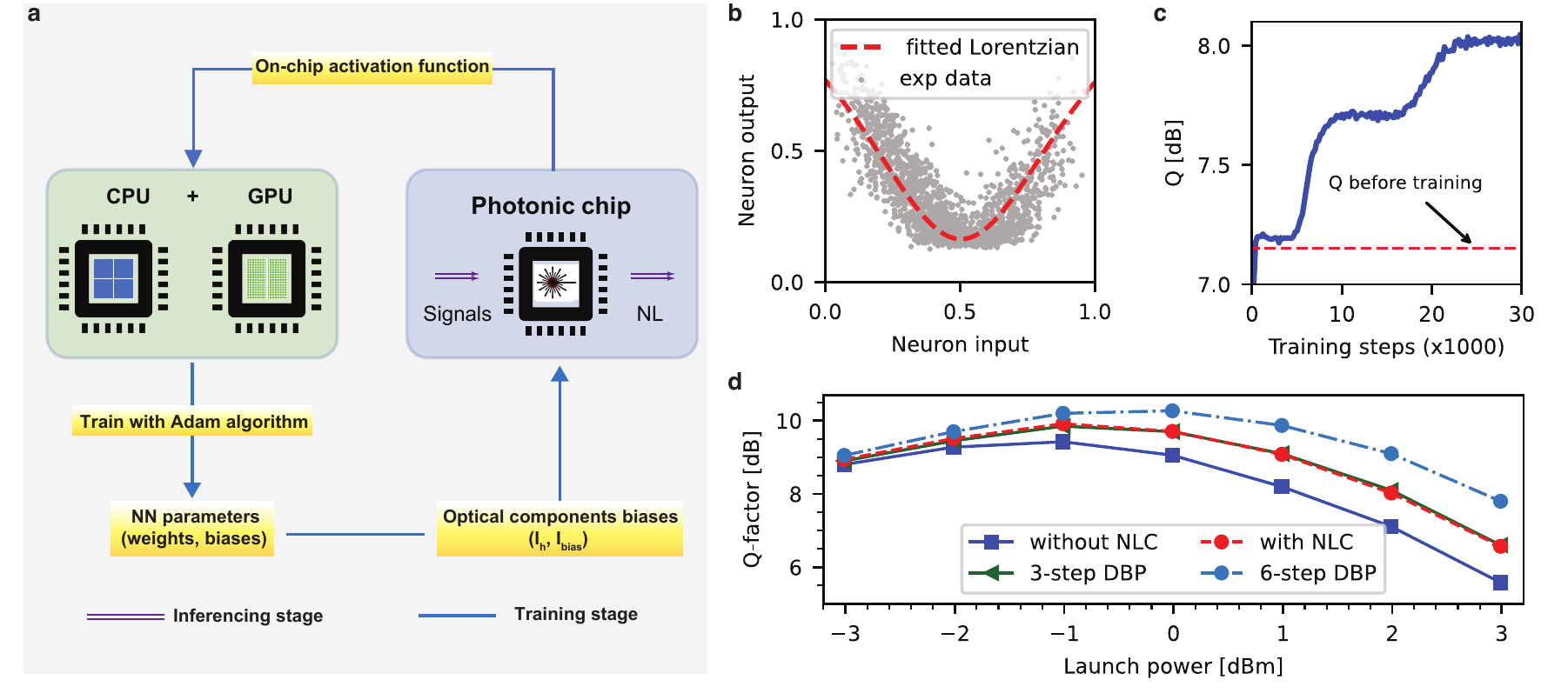}
\spacing{1.08}
\caption{\label{fig:training} (a) Flowchart of training process. (b) Activation function of photonic neuron. (c) The Q factor converges as the NN is trained, 32,016 symbols generated from the transmission system to monitor the training process; red trace: Q factor before nonlinearity compensation. (d) Transmission performance comparison. The neural network model for NLC is trained with the activation function of photonic neuron. The 3-step and 6-step DBP mean that fiber nonlinearity is compensated based on the conventional DSP approach digital backpropagation by equally dividing 10,080 km transmission link into 3 and 6 sections, respectively.}
\end{figure}


\section{Training with photonic neural network} 
The proposed PNN provides trainable parameters including both weights and biases. Trainable biases are critical for the NLC application, because they can optimize the transfer function shapes to best approximate the higher-order nonlinearity features from the symbols. To ensure that the trained NN parameters can be accurately mapped on the photonic device, we include the activation function of photonic neurons in the training phase. As shown in Fig.~\ref{fig:training}(a), we begin by characterizing the on-chip activation function by sending a training data sequence into the weight bank and recording the corresponding neuron output waveform. The transfer function is fitted to a Lorentzian function, as shown in Fig.~\ref{fig:training}(b), since the E/O response of MRR modulators is typically a Lorentzian function. The training data have the same data rate as the test data, in order to capture the bandwidth and the noise in the device that would affect the activation function shape. Following that, we apply the characterized activation function in the NN model and train the neural network with the Adam learning algorithm\cite{kingma2014adam}. 


The NN is trained with 32,106 training symbols in a single channel 32 Gbaud PM-16-QAM signal transmitted over a 10,080 km pure silica core fiber link. The weights are constrained to a range of -1 to 1 during the training, according to the operation range of the MRR weight bank. We monitor the Q-factor of a cross-validation (CV) data set to ensure that the NN is successfully trained to recognize the fiber nonlinear distortion. The CV data set contains 32,106 symbols and is independent with the training data set. As shown in Fig.~\ref{fig:training}(c), the Q-factor of the CV data increases gradually in the training process and finally converges to 8.1 dB after 21,600 steps. To enable efficient training, here we train the NN with a launch power higher than the optimal value. After training, a test data set with varying launch powers is sent to the optical link and processed by the trained NN. The nonlinear perturbation obtained from the NN is weighted by the difference in the launch power between the training and test set, and is then subtracted from the received signal. The Q factor of the signal after NLC is calculated and plotted in Fig.~\ref{fig:training}(d). The optical power at the Q factor peak is the optimal power that should be launched in the transmission fiber. Using the simulated PNN, the Q factor of the transmission link is increased by 0.66 dB as compared to that without NLC. Compared with 3-step DBP, the PNN is capable of achieving comparable Q factor performance.

\begin{figure}
\includegraphics[width=1\linewidth]{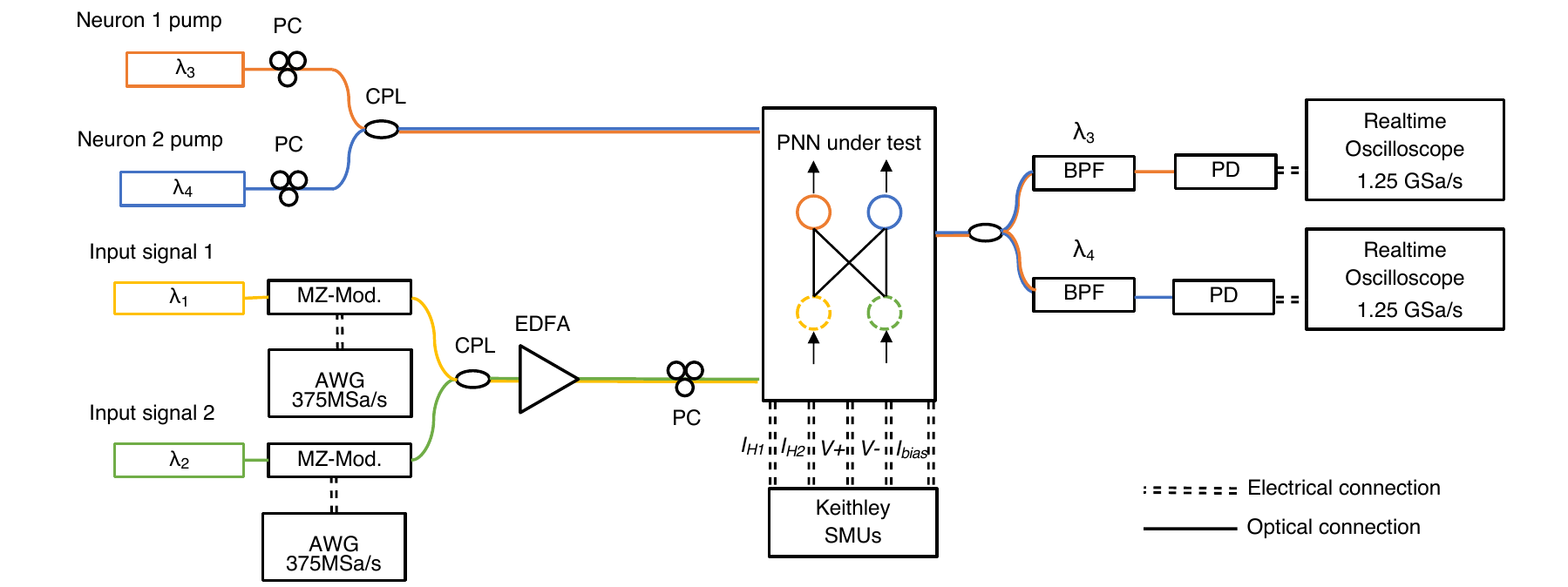}
\spacing{1.08}
\caption{\label{fig:setup} Experimental Setup. MZ-Mod: Mach-Zehnder modulator; AWG: arbitrary waveform generator, CPL: coupler; EDFA: erbium-doped fiber amplifier, PC: polarization controller; SMU: source meter unites; PNN: photonic neural network; PD: photodetector.
}
\end{figure}


\section{Photonic neural network implementation}
Following training, we implement the trained NN model to the PNN device, and evaluate whether the PNN can accurately compensate fiber nonlinearity impairment for high-speed signals. (See Supplementary Table.~3 for the weights and biases values). The experimental setup is shown in Fig.~\ref{fig:setup}. To emulate the first hidden layer outputs, two external cavity lasers are modulated with two waveforms generated from a two-channel arbitrary waveform generator (AWG). Each channel operates at a sampling rate of 375 MSample/s and generates a waveform of 32,106 test symbols at a symbol rate of 46.875 Mbaud/s. The test signal is independent of the signals in the training phase. The wavelengths of two lasers are initially aligned with the resonances of two MRRs in the MRR weight banks. The two signals are then combined and coupled into the PNN chip via a grating coupler. Two neuron pumps are derived from another two external cavity lasers with wavelengths close to the resonances of the MRR modulators. The modulator’s p-n junction is forward biased, allowing photocurrents from the balanced photodetector to modulate the neuron pumps through free-carrier injection. Under the carrier injection mode, the measured neuron bandwidth is 150 MHz. In Supplementary information Fig.~2, we show that, with the same photonic circuits and optimized MRR modulator geometry, the neuron bandwidth can be practically improved to 10 GHz.

\begin{figure}
\includegraphics[width=1\linewidth]{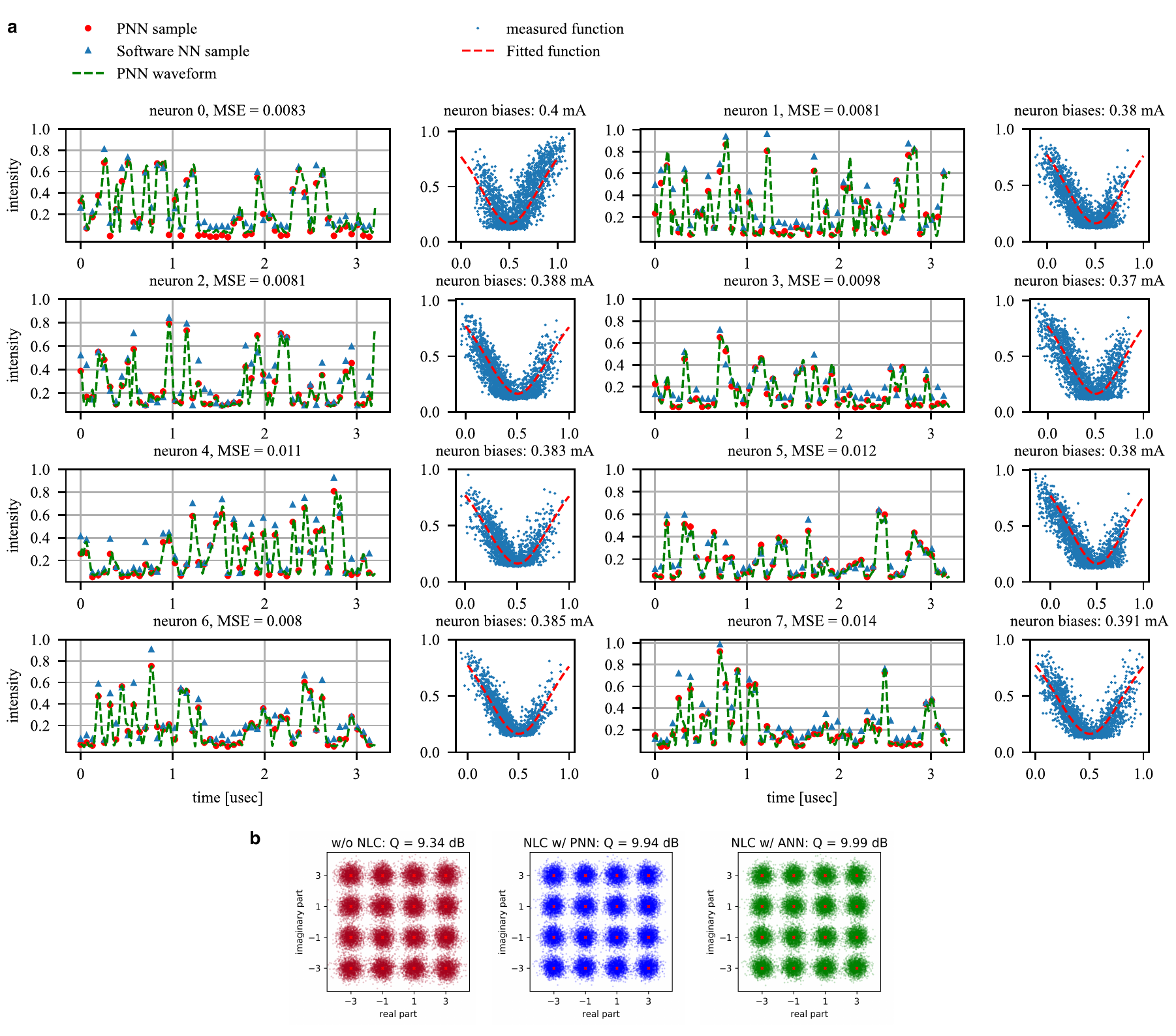}
\spacing{1.08}
\caption{\label{fig:results_2}(a) waveforms at the second hidden layer output. Green dashed line: photonic neuron waveform; red dot: photonic neuron waveform downsampled to the data rate of 16-QAM signals; blue triangle: simulated waveform after downsampling. Ideally, the red points should overlap with the blue. And biased activation function characterized from the measured waveforms (blue dot) versus ideal activation function. (b) constellations of the 16-QAM signal on the X-polarization. Red: without nonlinearity compensation (NLC); blue: NLC using PNN; green: NLC using simulated NN. Red cross: constellation diagram of 16QAM without noise.}

\end{figure}

The trained NN parameters are grouped into four subsets, each with a 2$\times$2 weight matrix and 2 neuron biases.  Each weight matrix is mapped to a current matrix $\bm{I_h}$ and applied to the in-ring heaters of the MRR weight banks using the control method described in Ref~\cite{huang2020demonstration}. To apply neuron biases, the modulator bias current $I_{bias}$ is tuned to approximate the trained transfer function. The 2$\times$2 current matrix and 2 modulator biases are simultaneously applied to the PNN chip. Fig.~\ref{fig:results_2}(a) shows the waveform snapshots from the PNN output, which presents the outputs of the second hidden layer, together with the measured transfer function of the eight neurons. Each neuron's transfer function corresponds to the different piece of the characterized Lorentzian function under intended neuron biases. We simulate the same NN with a 32-bit GPU-assisted workstation to benchmark the accuracy of PNN implementation. Each neuron is evaluated by comparing the experimental PNN output (red dots) with the simulated NN (blue dots). The mean squared errors (MSEs) of photonic neurons are calculated over 32,016 test symbols. MSEs are used to quantify the errors caused by noises and derivations from target weights and activation functions. The MSEs vary between 0.008 to 0.014. The signal's Q factor is then calculated from the eight photonic neuron outputs (see Supplementary information \textit{Q factor calculation}). Without nonlinearity compensation, the optimal Q-factor of the signal is 9.34 dB, obtained at the launch power of -1 dBm. With the PNN, the signal's Q factor is improved by 0.60 dB, and the constellation diagram after NLC is plotted in Fig.~\ref{fig:results_2}(b.i), showing less noise in contrast to that without NLC. The Q-factor obtained from the simulated NN indicates a 0.65 dB gain in Q-factor. That is, the penalty of loading the neural network to the PNN is only 0.05 dB after taking into account all inherent noise from the physical components and equipment.

\section{Scalability for NLC}
Mainstream silicon photonic platforms offer an opportunity to implement the entire neural network for NLC on a single silicon photonic chip. Silicon photonic circuits with a few thousands of elements have been demonstrated~\cite{sun2013large,poulton20208192}. Weight pruning\cite{zhang2019field} is an effective pathway for reducing the PNN footprint and power consumption, which can be accomplished by having a few more neurons in hidden layers (see Fig.~\ref{fig:activation_function_comparison}) while reducing the number of input triplets. The current NN model has 892 input triplets. After weight pruning, the number of inputs can be reduced by three times to 300. The photonic neural network, in this case, has 952 optoelectronic components. These components require 0.38 mm\textsuperscript{2} chip space. Electrical traces and pads connecting to the optoelectronic components provide electrical biases to the photonic circuit.
Electrical traces and pads connecting to the optoelectronic components provide electrical biases to the photonic circuit. The amount of chip space used by electrical traces and pads varies significantly depending on the chip floorplan and packaging technologies. For example, the electrical traces can be very short using the flip-chip bonding technology, and the pads take up 0.76 mm\textsuperscript{2} areas (each bonding pad takes 20$\mu$m $\times$ 20$\mu$m area)\cite{carroll2016photonic}. Ref~\cite{sun2013large} have demonstrated that silicon photonic devices with comparable footprints are feasible.

A challenge of large-scale silicon photonic circuits is device non-uniformity, i.e., the dimensions and performances of the devices across the chip are not uniform and deviate from the design. In PNNs, device non-uniformity will cause variations of the relationship between the actuated MRR currents and the actual weights, the activation function shape, etc. The influences of device variations can be minimized by pre-calibrating each device or using control methods, such as the feedback MRR control~\cite{tait2018feedback,huang2020demonstration}. Device variances can be further reduced through post-fabrication technologies such as trimming~\cite{atabaki2013accurate} or in-situ training\cite{wang2019situ}. Another challenge deals with light sources. Frequency combs-based WDM sources\cite{brasch2016photonic} allow the generation of broadband and evenly spaced emission wavelengths from a single laser, and thus offer a practical approach of generating WDM light sources for the PNN. Soliton-based frequency combs built on CMOS-compatible photonic integrated circuits can generate hundreds of comb lines. Such systems can integrate with the PNN device. 

\section{Claiming cascadability for NLC}

We discuss the feasibility of scaling the current chip to a multi-layer neural network. This issue refers to the need for cascadability, that is, the ability to have the one neuron output to excite the neurons in the next layer. Cascadability requires 1) the neuron output can evoke an equivalent response in neurons in the subsquent layer; 2) to avoid noise propagation along with multiple neural layers. The main conclusion is, in order to have one layer driving the next layer, enough power must be provided to compensate for loss and power splitting due to fan-out to the next layer. Such power can come from optical lasers or electrical gain via a high transimpedance. The electrical gain can be provided with a passive resistor or an active TIA, both of which are compatible with today’s silicon photonic platform. For example, we have implemented an on-chip resistor in our current PNN design. Alternatively, we can use an efficient modulator (i.e. small $V_{\pi}$ and junction capacitor C) to reduce the power requirement. We derive the devices and power requirement of maintaining the cascadability in the Supplementary information \textit{
Performance comparison between DSP and PNNs}.

Regarding noise propagation, we found that if the voltage swing to the modulator is large enough, the nonlinear transfer function of the modulator can suppress the noise and avoid noise propagation across the network\cite{de2019noise}. Such requirements can also be addressed with sufficient power or with an efficient modulator, as argued in~\cite{nozaki2019femtofarad,de2019noise,nahmias2019photonic}.

\section{Impact of activation function and the number of neurons}

\begin{figure}[ht]
\center\includegraphics[width=0.5\linewidth]{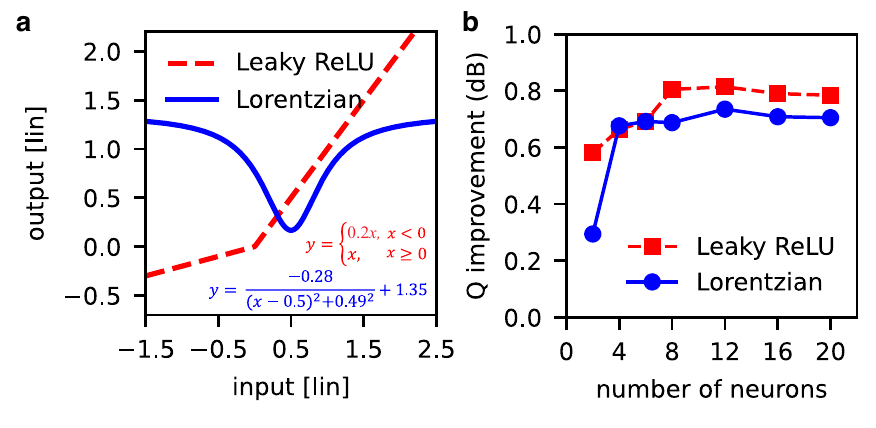}
\caption{\label{fig:activation_function_comparison}(a)~Leaky ReLU, optimized for fiber nonlinearity compensation as demonstrated by Zhang et al.\cite{zhang2019field}, and Lorentzian, characterized from the PNN. (b)~The comparison of Q factor improvement between Leaky ReLU (red) and Lorentzian (blue) activation functions.} 
\end{figure}

In the NLC application, the NN is trained to recognize the fiber nonlinearity perturbation from the received symbols. An appropriate choice of activation function can benefit such regression problems by including higher-order nonlinearities among the symbols. Fiber nonlinearity perturbation manifests itself as long, heavy-tailed symbols due to nonlinear phase noise. To extract such features, especially the tail edge, an activation function with an unbounded range is preferred. Fig.~\ref{fig:activation_function_comparison}(a) shows the performance gap between the Lorentzian function of an MRR modulator neuron and the Leaky ReLU function which has no upper boundary for positive input. Generally, for different NN applications, activation functions need to be chosen to synthesize particular tasks. In contrast to a single cavity MRR, photonic devices with multiple coupled cavities can exhibit different lineshapes beyond the Lorentzian shape. Additional transfer function programmability could be achieved at the expense of additional circuits and calibration complexity using multiple-cavity modulators for some or all of the neurons\cite{huang2020chip}. Reconfigurable elements such as heaters or phase change materials allow optimizing device's transfer functions for different machine learning tasks. 

In addition, as shown in Fig.~\ref{fig:activation_function_comparison}(b), the Q-factor improves as the number of neurons on the second-hidden layer increases. This can be expected because, with more neurons, the network can more precisely constitute different parts of nonlinearity perturbation over symbols. PNNs add almost no delays as the NN size increases, because the major operations in a PNN are computed at a single time step.


\subsection{}

\section{Conclusion}

We have reported a reconfigurable photonic-electronic integrated neural network platform that incorporates photonic components on a silicon photonic platform for high-speed information processing. We demonstrated our system is capable of performing essential functionalities such as accurate weighting, summation, and biased nonlinearity for neural network computing. We applied this system for fibre nonlinearity compensation in trans-Pacific transmission data links, which require extremely high throughput and real-time signal processing. We experimentally implemented the second layer of the neural network using silicon photonic neurons, and showed in simulations that the extended system can accurately model the fibre nonlinearity, leading to a Q-factor improvement comparable with numerical simulations on a conventional computer. The system is capable of processing optical signals in the analogue domain, thereby significantly relaxing the stringent requirements of complexity and speed that are typically placed on conventional DSP circuits. Our approach could help in the development of reconfigurable photonic–electronic integrated circuits for solving a range of problems in high-speed intelligent signal processing.

\begin{methods}

\subsection{Device Fabrication}

The device sample is fabricated on a silicon-on-insulator (SOI) wafer with a silicon thickness of 220 nm and a buried oxide thickness of 2 $\mu$m. The waveguide is 500 nm wide. The weight bank consists of four MRRs coupled with two bus waveguides in an add/drop configuration. The four MRRs have radii of 8.0, 8.1, 8.2, 8.3 $\mu$m, respectively. Only the first two MRRs are used in the experiments. A slight difference is introduced in the ring radii to avoid resonance collision. The gap between the ring and bus waveguide is 200 nm, yielding a Q factor of $\sim$6000. For the purpose of weight control, in-ring N-doped photoconductive heaters are used. To implement the N-doped heater, each MRR consists of a ring waveguide etched to a 90 nm thick pedestal that hosts the phosphorus dopants. A 10 $\mu$m wide N doping section is patterned to follow the MRR, outside of which heavy N++ doping is used to make ohmic contacts. Metal vias and traces are deposited to connect the heater contacts of the MRR weight bank to electrical metal pads.

The MRR modulator consists of a ring waveguide with a radius of 8 $\mu$m coupled to two bus waveguides with gaps of 0.2 and 0.5 $\mu$m respectively. The ring waveguide is etched to a 90-nm-thick pedestal that hosts boron and phosphorus to form n+ and n-, and p+ and p- regions, respectively. The lightly p- and n-doped regions are extended to the waveguide core to form a p-n junction for high-speed modulation. The highly p+ or n+ doped regions are located at the pedestal edge to serve as p-type and n-type ohmic contact areas. To fabricate the germanium-on-silicon photodetector, a germanium layer is deposited and patterned on top of the silicon waveguide. Boron and phosphorus are implanted in the sidewalls and slabs of the Ge waveguide to form a horizontal p-i-n junction as well as p-type and n-type ohmic contact areas. Metal vias and traces are deposited to connect the electrical ports of the devices to the electrical metal pads. The cross-sections of MRR resonator embedded with n-doped heater, MRR modulator, and germanium-on-silicon photodetector are illustrated in Supplementary Fig.~1.

The on-chip capacitor is formed by two layers of metal and has a capacitance of 2 pF. The resistor is formed by a slab waveguide etched to a 90 nm thick pedestal that heavily doped with phosphorus (N++). The resistor has a resistance of 300 $\Omega$.

\end{methods}




\begin{addendum}

 \item [Data availability] All data used in this study are available from the corresponding author upon reasonable request.
 \item [Code availability] All codes used in this study are available from the corresponding author upon reasonable request.
 
\end{addendum}

\begin{addendum}
 \item This research is supported by the Office of Naval Research (ONR) (N00014-18-1-2297), Defense Advanced Research
Projects Agency (HR00111990049), National Science Foundation (NSF) (Grants No. ECCS 1642962 and No. DGE 1148900), and CUHK Research Direct Grant. The devices were fabricated at the IME ASTAR foundry in Singapore. Fabrication support was provided via the Natural Sciences and Engineering Research Council of Canada (NSERC) Silicon Electronic-Photonic Integrated Circuits (SiEPIC) Program and the Canadian Microelectronics Corporation (CMC). 
 
 \item[Author Contributions]  C.H., S.F., and T.L. conceived the ideas and implement the experimental setup, designed the experiment, conducted the experiment measurement, and analyzed the results. C.H., T.L., E.B., S.B., A.J., H.P. designed the silicon photonic chip. A.T., Y.T., F.Y, B.S. provided the theoretical support. T.L., E.B., S.B. performed the chip packaging. C.H., S.F., T.L., A.T., B.S. wrote the manuscript. T.W. and P.P. supervised the research and contributed to the general concept and interpretation of the results. All authors discussed the data and contributed to the manuscript.
 \item[Competing Interests] The authors declare no competing interests.
\item[corresponding authors ] Chaoran Huang~(email: chaoranh@princeton.edu, ORCID: 0000-0001-6997-758X), Paul R. Prucnal~(email: prucnal@princeton.edu, ORCID: 0000-0002-5291-5830).
\end{addendum}

\newpage

\subsection{References}

\bibliography{ref}


\newpage

\end{document}